%% file: actor-based-responsibility.tex
\documentclass[envcountsect,envcountsame]{llncs}
\usepackage[maxbibnames=99]{biblatex}
\usepackage{xspace}
\input{packages.tex}

\newcommand{\aplayser}{$a$\xspace}

\newcommand{\acktext}{
The authors are supported by the DFG through
        the DFG grant 389792660 as part of TRR~248 (see \protect\url{https://perspicuous-computing.science})
        and the Cluster of Excellence EXC 2050/1 (CeTI, project ID 390696704, as part of Germany's Excellence Strategy)
        and by BMBF (Federal Ministry of Education and Research) in DAAD project 57616814 (SECAI, School of Embedded and Composite AI) as part of the program Konrad Zuse Schools of Excellence in Artificial Intelligence.
}

\title{Responsibility in Actor-Based Systems}
\author{Christel Baier\inst{1, 2} \and Sascha Klüppelholz\inst{1} \and Johannes Lehmann\inst{1,2} \thanks{Authors are listed in alphabetical order.} \thanks{\acktext}}

\institute{Technische Universität Dresden, Dresden, Germany
	\and Centre for Tactile Internet with Human-in-the-Loop (CeTI)}

\begin{document}
	
	\maketitle

\input{abstract}

    \input{new-introduction}
	\input{02-prelims}

\input{03-responsibility}
	\input{04-from-reactive-modules}
	\input{06-experiments}
	\input{07-conclusion}

		\emergencystretch=2em
	\printbibliography
\end{document}

%% file: packages.tex
\input{fix-counter}

\usepackage{mathtools}
\usepackage{amssymb}
\usepackage{stmaryrd}
\usepackage[dvipsnames]{xcolor}
\usepackage[color=SpringGreen]{todonotes}
\usepackage[altpo]{backnaur}
\usepackage{subfig}
\usepackage{makecell}
\usepackage{listings}
\usepackage{pgfplots}
\pgfplotsset{compat=1.15}
\usepackage{tikz}
\usetikzlibrary{automata,positioning,shapes.geometric}
\usepackage{hyperref}
\usepackage[hyphenbreaks]{breakurl}

\newcommand{\ltuple}{\left(}
\newcommand{\rtuple}{\right)}

\renewcommand{\ts}{\mathcal{T}}
\newcommand{\tsStates}{S} % states of transition system
\newcommand{\tsTrans}{\to} % transition relation of transition system
\newcommand{\tsInit}{s_{\mathit{init}}} % initial state of transition system
\newcommand{\tsBad}{\lightning} % the undesirable states
\newcommand{\tsTuple}[1][]{\ltuple \tsStates#1, \tsTrans#1, \tsInit#1, \tsBad#1 \rtuple}

\newcommand{\tsRun}{\rho} % run of transition system
\newcommand{\counterexample}{\hat{\rho}} % counter-example
\newcommand{\game}{\mathcal{G}}
\newcommand{\safe}{\textit{Safe}}
\newcommand{\reach}{\textit{Reach}}
\newcommand{\reachStates}{S_{\reach}}
\newcommand{\safeStates}{S_{\safe}}
\newcommand{\gameStates}{S}
\newcommand{\gameInit}{s_{\textit{init}}}
\newcommand{\gameTrans}{\to}
\newcommand{\gameBad}{\lightning}

\newcommand{\gamePlay}{\rho}
\newcommand{\winningRegion}{\mathit{Win}}

\newcommand{\gameVal}{\mathit{val}}
\newcommand{\auxHelp}{S_\mathit{Aux}}
\newcommand{\auxAdv}{S_\mathit{Adv}}
\newcommand{\actorSet}{\ltuple \actors, \auxHelp, \auxAdv \rtuple}
\newcommand{\actors}{\mathcal{A}}
\newcommand{\Actors}{\mathcal{A}}
\newcommand{\actor}{a}

\newcommand{\coopGame}{\gamma}
\newcommand{\shapley}[1]{\mathit{Shap}_{#1}}
\newcommand{\tsToGameFw}{\textsf{GameFw}}
\newcommand{\tsToGameBw}{\textsf{GameBw}}

\newcommand{\tsCoopGameFw}[1]{\textsf{CoopFw}_{#1}}
\newcommand{\tsCoopGameBw}[2]{\textsf{CoopBw}^{#2}_{#1}}
\newcommand{\tsCoopGame}[2]{\textsf{Coop}}

\newcommand{\act}{\alpha}

\newcommand{\acts}{\mathit{Act}}

\newcommand{\safetyFormula}{\varphi_{\lightning}}

\newcommand{\scheduler}{\mathsf{WithSched}}
\newcommand{\actionSeparated}{\mathsf{ActionSeparated}}

%% file: abstract.tex
	\begin{abstract}
          The enormous growth of the complexity of modern computer systems
          leads to an increasing demand for techniques that support the 
          comprehensibility of systems. This has motivated the very active research
          field of formal methods that
          enhance the understanding of
          \emph{why} systems behave the way they do.
          One important line of research within the verification community relies on
          formal notions that measure the degree of responsibility
          of different actors.
          In this paper, we first provide a uniform presentation of recent
          work on responsibility notions based on Shapley values
          for reactive systems modeled by  
          transition systems and considering safety properties.
          The paper then discusses how to use these formal 
          responsibility notions 
          and corresponding algorithms for three different types of actor sets:
          the {\emph{module-based notion}} serves to reason 
          about the impact of system components on the satisfaction
          or violation of a safety property. Responsibility values for 
          {\emph{value-based actor sets}} and {\emph{action-based actors}} 
          allow for the identification of program instructions and control 
          points that have the most influence on a specification violation.
          Beyond the theoretical considerations, this paper reports on experimental 
          results that provide initial insights into applicability and scalability.
	\end{abstract}

%% file: new-introduction.tex
\section{Introduction}

Ensuring the correctness of software is of ever-increasing importance.
This has given rise to a variety of formal techniques that aim to
prove that software adheres to its specification. 
The classical verification task is to determine whether a model
satisfies a logical formula. Algorithmic techniques such as model
checking \cite{CPG-MCBook,BK-MCBook,Handbook-MC} can solve this task
for systems represented by a finite-state model and temporal logical
specifications. Moreover, if a model checker finds a specification
violation, it can return a \emph{counterexample}, which is an
execution sequence for which the specification does not hold. Some
verification techniques also support the generation of a mathematical
\emph{certificate} in case that the specification holds. 
Such certificates offer the possibility to check the verification result
independently from the verification process. Examples are inductive
variants, derivations in deductive systems or mathematical proofs 
\cite{Namjoshi-certifyingMC,Temporal-Safety-Proofs-CAV02,Leroy-backend,Necula-proof-carrying-code,Necula-et-al-VMCA11,certificates-mu-calculus-SPIN16}.

Although such certificates and counterexamples can provide useful
insights, they are often complex and can enhance the comprehensibility
of the system behaviors only to a very limited extent. This motivated
research on additional techniques that provide comprehensible
and more coherent explanations for \emph{why} a system behaves the way it does. 
Most notable is the work on methods that extract additional debugging 
information from counterexamples
\cite{ErrorsInCounterexamples,ErrorDistanceMetrics,
ExplainingCounterexamples,Beer-et-al-Explaining-counterexamples,
StatisticalFaultLocation-FASE15,SPINCause,BackwardResponsibility}. 
The majority of this work is based on formal notions of
\emph{counterfactual causality} and inspired by Halpern and Pearl's
pioneering work on actual causality for structural equation models
\cite{HalpernPearlI, HalpernPearlII, HalpernPearlModification}. The
main idea of the counterfactual principle for cause-effect relations
is that the effect would not have happened if the cause had not
occurred. Related to these works on causality-based explanations of
counterexamples is research on formal notions to quantify the 
\emph{(degree of) responsibility}. Such quantitative notions aim to 
measure the individual influence of actors of a system on (potential 
or actually observed) effects on the systems behavior.

Research on responsibility notions in the verification context is
motivated by natural questions. For instance, if a system fails to
meet its specification, which system component is most responsible? 
Or, which lines of program code are erroneous and lead to the
specification violation? Such questions naturally arise 
from a \emph{backward-looking} perspective where an execution sequence with
undesired behavior has been observed or has been reported in the form
of a counterexample by a model checker. Opposed to this is the
\emph{forward-looking} perspective where all potential executions of the
system are taken into account and one asks for responsibilities for
all potential specification violations (or, conversely, the satisfaction 
of the specification if the system meets its specification).
The forward-looking perspective is not only useful for classical functional
specifications from the safety-liveness spectrum, but also for
reasoning about non-functional properties, e.g., by asking for the
system components that have most impact on the expected energy
consumption.
The distinction between the forward- and backward-looking perspective goes to
van de Poel \cite{ForwardBackward} who studied responsibility under
philosophical aspects. Among others, it has been taken up for
multi-agent systems and game structures under partial observability
assumptions \cite{Yazdanpanah-strategic-resp-2019,GameTheoreticResponsibility}.

Orthogonal to the distinction between forward- and backward-looking approaches 
are the different possibilities to measure the individual impact of
actors on effects. The most popular concepts that have been used in
the verification context are \emph{intervention-based approaches}
(e.g. those used in \cite{Chockler-et-al-STE-ref-TACAS08,What-causes-a-system-ToCL08,Chockler-caus-resp-CREST16}) 
and more recent approaches that rely on \emph{Shapley values} (e.g. those used in
\cite{Resp-paramMC-AAAI21,GameTheoreticResponsibility,ForwardResponsibility,BackwardResponsibility}). 
Responsibility notions of the intervention-based approaches rely on the general concept of
responsibility introduced by Chockler and Halpern
\cite{CH-degree-of-resp} and derive the degree of responsibility from
the smallest number of changes to the system model (``interventions'')
that make an event a cause of another one with respect to Halpern and
Pearl's notion of causality. On the other hand, Shapley values
\cite{ShapleyValue} are a game-theoretic concept that has been
introduced for one-shot cooperative games which assign a numerical game
value to any coalition (i.e., subset) of players. These values
represent the outcome the players of a coalition can achieve when
cooperating with each other. Roughly speaking, the Shapley value of a
player \aplayser measures the individual impact of \aplayser by its average
contribution to the outcomes of coalitions that include player \aplayser. 
The idea of using Shapley values in the verification context is to assign to
each coalition of actors of a system (e.g., the system components) the
game value 1 if these actors can enforce the specification by
resolving their nondeterministic choices in an appropriate way, and 0
otherwise. For other types of responsibility in AI systems and
causality-based explanations in the verification context, see, e.g., 
the survey articles \cite{Chockler-caus-resp-CREST16,Responsibility-in-AI-survey,From-Verification-to-Explication-ICALP21}.

This article focuses on responsibility notions defined by Shapley
values. More concretely, we present the notions of forward
\cite{ForwardResponsibility} and backward responsibility
\cite{BackwardResponsibility} of \emph{actors} in transition systems on
the satisfaction or violation of safety properties in a uniform way.
In this setting, the actors are logical units of a system, formalised
here by sets of states that constitute a partition of the state
space or a fragment thereof. The idea is that each actor controls the
nondeterministic choices in the states assigned to her. Beyond the
works \cite{ForwardResponsibility,BackwardResponsibility} that
concentrate on the theoretical foundations, we discuss the generation
of natural actor sets, which are either {\emph{module-based}}, 
{\emph{value-based}}, or {\emph{action-based}}. The proposed techniques 
for generating actor sets operate on the syntactic description of the system 
by a program code written in a \emph{reactive module language}. 
Value-based actors are determined by partial assignments for the program's
variables and can, for instance, serve to identify the control points
or instructions of a program that have the most influence on a
specification violation. Our techniques for module-based actors 
and action-based actors rely on model transformations. 
The reason is that each state of a transition system can have transitions 
of multiple modules. Thus, modules cannot be treated as actors directly. 
Here we propose a model transformation that generates a new program with 
an auxiliary scheduler module that serves to split such states where two or 
more modules can make a move. Likewise, reasoning about the responsibility
of actions requires a tailored model transformation.

The remainder of this paper is structured as follows.
Section~\ref{sec:prelims} presents our notations for transition
systems, safety games and Shapley values. Section~\ref{sec:actors}
provides a summary of the results presented  in
\cite{ForwardResponsibility} and \cite{BackwardResponsibility} for
safety properties (rephrased in a uniform manner). In
Section~\ref{sec:reactive_modules}, we show how actors can be derived
from a reactive module specification.
Section~\ref{sec:implementation} reports on experimental studies that
have been carried out a prototypical implementation and provide
insights on how the proposed responsibility notions can be used to
locate errors in system models. Section~\ref{sec:conclusion} concludes
with a brief summary and a discussion on future research directions.

%% file: 02-prelims.tex
\section{Preliminaries}

\label{sec:prelims} \label{sec:prelim}

We briefly present the notations for transition systems, safety games
and Shapley values. More details about transition systems and (safety)
games can be found for example in the text books
\cite{CPG-MCBook,BK-MCBook,Games-Book-GTW,Games-Book-Nath-et-al}.

\paragraph*{\bf Transition systems.} A \textit{transition system} is a
tuple $\ts = \tsTuple$ where $\tsStates$ is a finite set of 
\textit{states}, $\tsTrans \, \subseteq \, \tsStates \times \tsStates$
the \emph{transition relation} on $\tsStates$, $\tsInit \in \tsStates$
the \textit{initial state}, and $\tsBad \subseteq \tsStates$ a set of
\emph{bad states}. We assume that the states in $\tsBad$ are
absorbing. A \textit{run} on $\ts$ is an infinite sequence of states
$\tsRun = \tsRun_0 \tsRun_1 \ldots \in \tsStates^\omega$, where
$\rho_0 = \tsInit$ and for all $i \in \mathbb N$, we have $\tsRun_i
\tsTrans \tsRun_{i+1}$. The states in $\tsBad$ induce a safety
property ``never reach $\tsBad$'', which is satisfied for a run
$\tsRun = \tsRun_0 \tsRun_1 \ldots \in \tsStates^\omega$ if
$\tsRun_i\cap\tsBad =\varnothing$ for all $i\in\mathbb{N}$ and
satisfied for $\ts$ if all runs satisfy the safety property.

We call $\counterexample = \counterexample_0 \ldots \counterexample_k
\in \tsStates^\ast$ a counterexample to the safety property ``never
reach $\tsBad$'' if $\counterexample$ is the prefix of a run with
$\counterexample_k \in \tsBad$. Additionally, we may assume that
$\counterexample$ is loop-free, i.e. $\counterexample_i \neq \counterexample_j$ for all $i \neq j$.

\paragraph*{\bf Safety games.} A \emph{safety game} is a tuple $\game
= \ltuple \safeStates, \reachStates, \gameTrans, \gameInit, \gameBad
\rtuple$, where $\safeStates$ and $\reachStates$ are disjoint sets of
states such that the states in $\safeStates$ are controlled by player
$\safe$, while the states in $\reachStates$ are controlled by player
$\reach$, $\gameTrans$ is the \emph{transition relation} on
$\gameStates = \safeStates \cup \reachStates$, $\gameInit \in
\gameStates$ is the \emph{initial state} and $\gameBad \subseteq
\gameStates$ is the set of \emph{bad states}. 
A \emph{play} $\gamePlay \in \gameStates^\omega$ is an infinite
sequence $\gamePlay_0 \gamePlay_1 \ldots$ such that $\gamePlay_0 =
\gameInit$ and $(\gamePlay_i, \gamePlay_{i+1}) \in \, \gameTrans$ for
all $i \in \mathbb N$. A play $\gamePlay$ is \emph{winning} for
$\reach$ if $\gamePlay_i \in \gameBad$ for some $i \in \mathbb{N}$,
otherwise it is winning for $\safe$. A strategy for $\safe$ is a
function $\sigma \colon \safeStates \gameTrans \gameStates$ with $(s,
\sigma(s)) \in \, \gameTrans$ for all $s \in \safeStates$ (and
strategies for $\reach$ are defined similarly). A pair of strategies
$\ltuple \sigma_{\safe}, \sigma_{\reach} \rtuple$ for $\safe$ and
$\reach$ induces a play $\gamePlay = \gamePlay_0 \gamePlay_1 \ldots$
with $\gamePlay_{i+1} = \sigma_{\safe}(\gamePlay_i)$ if $\gamePlay_i
\in \safeStates$ and $\gamePlay_{i+1} =
\sigma_{\reachStates}(\gamePlay_i)$ otherwise. A strategy for $\safe$
is winning if, for all strategies of $\reach$, the induced play is
winning for $\safe$ (and winning strategies for $\reach$ are defined
similarly). For any game $\game$, we write $\gameVal(\game) = 1$ if
there exists a winning strategy for $\safe$ and $\gameVal(\game) = 0$
otherwise.
The winning region $\winningRegion(\game)$ contains all states from
which $\safe$ has a winning strategy. Computing the winning region takes linear time in the number of states and transitions of the model. \cite[Theorem~12]{Games-Book-Nath-et-al}.

\paragraph*{\bf Shapley values.} Shapley values \cite{ShapleyValue}
are a prominent concept to measure the individual impact of agents on
the outcome of plays in a game. They have been introduced for
cooperative games with real-valued outcomes. For the purposes of this
paper, simple cooperative games with monotonic Boolean functions for
the possible outcomes are sufficient.

Let $\actors = \{a_1, \ldots, a_n\}$ be a set of players. A
\emph{simple cooperative game} is a monotonic function $\coopGame
\colon 2^{\actors} \mapsto \{0, 1\}$.
For a simple cooperative game $\coopGame$, the Shapley values are
defined by the function $\shapley{\coopGame} \colon \actors \to [0,
1]$ given by \[\shapley{\coopGame}(a) \ = \sum_{C \subseteq \actors
\setminus \{a\}} \frac{(|\actors| - |C| - 1)! \cdot |C|!}{|\actors|!}
\Big(\coopGame(C \cup \{a\}) - \coopGame(C)\Big). \]

For $\actor \in \actors$ and $C \subseteq \actors \setminus \{ a \}$,
we call $\ltuple C, \actor \rtuple$ a \emph{switching pair} if
$\coopGame(C \cup \{a\}) - \coopGame(C) = 1$, i.e., if $\coopGame(C
\cup \{a\}) =1$ and $\coopGame(C) = 0$.

%% file: 03-responsibility.tex
\section{Forward and backward responsibility} \label{sec:actors}

Following previous works on notions for (the degree of) responsibility
in operational models
\cite{ForwardResponsibility,BackwardResponsibility,GameTheoreticResponsibility}, 
we present definitions for the forward and backward responsibility of \emph{actors} 
on the satisfaction (and the violation, respectively) of a safety property. 
The idea is that every actor controls a set of states and resolves the non-deterministic 
choices within these states. Moreover, there can be other states that are not assigned to
any actor. These can be auxiliary states that have been introduced for
modelling purposes or states whose behavior is given by an adversarial
environment (see Section \ref{subsec:action_based}). The
responsibility notions will then assign numerical values to the actors.

\begin{definition}[Actor] Let $\ts = \tsTuple$ be a transition system.
A \emph{responsibility signature} is a triple
$(\actors,\auxHelp,\auxAdv)$ where $\auxHelp$ and $\auxAdv$ are
disjoint subsets of the state space $S$ and $\actors$ a partition of
$S \setminus (\auxHelp \cup \auxAdv)$. The elements of $\actors$ are
called \emph{actors}. The states in $\auxHelp$ are called 
\emph{auxiliary states} and the states in $\auxAdv$ the
\emph{adversarial states}. \end{definition}

In Section~\ref{sec:reactive_modules}, we propose three schemes for
extracting actor sets from a reactive module specification. First,
actors can be extracted from modules, such that each module is
represented by an actor (Section~\ref{subsec:module_based}). In
Section~\ref{subsec:value_based}, value-based actors are constructed,
where each actor corresponds to a partial variable assignment. For
both module- and value-based actors. $\auxHelp$ and $\auxAdv$ will be
empty. An instance where $\auxHelp$ and $\auxAdv$ are nonempty are
action-based actors (Section~\ref{subsec:action_based}). For these,
every actor represents one action of the program.

For the remainder of this section, we suppose a fixed responsibility
signature $(\actors,\auxHelp,\auxAdv)$ and define the (degree of)
responsibility for the actors $a \in \actors$.

As in
\cite{ForwardResponsibility,BackwardResponsibility,
GameTheoreticResponsibility}, these notions rely on Shapley values for
one-shot cooperative games given by a function $\coopGame \colon
2^\actors \to \{0, 1\}$ that assign to every coalition of agents an
outcome. Intuitively, the outcome $\coopGame(C)=1$ for a coalition $C
\subseteq \actors$ indicates that the actors in $C$ can cooperate to
prevent the violation of the safety property, while $\coopGame(C)=0$
means that the actors in $A$ do not have such a strategy.

\begin{definition}[Safety games for coalitions of agents -- forward]
\label{def:ts_to_safety_fw} Let $\ts = \tsTuple$ be a transition
system, $\actorSet$ a responsibility signature and $C \subseteq
\actors$. We define the game \[\tsToGameFw(\ts, \auxHelp, C) = \ltuple
C \cup \auxHelp, \tsStates \setminus (C \cup \auxHelp), \tsTrans,
\tsInit, \tsBad \rtuple.\] \end{definition}

This construction preserves the structure of the transition system.
State ownership is assigned based on whether they are in the
coalition. Additionally, Player $\safe$ controls all auxiliary states.
If Player $\safe$ wins, this indicates that the states in $C$ and the
auxiliary states are sufficient to enforce the safety property.

As the function name indicates, this construction is tailored for
forward responsibility. For backward responsibility we incorporate 
the a given counterexample~$\counterexample$ into the construction.

\newcommand{\gameTransBw}{\gameTrans_{\mathit{Bw}}}
\begin{definition}[Safety games for coalitions of agents -- backward]
\label{def:ts_to_safety_bw} Let $\ts$, $\actorSet$ and $C$ be as in
Definition~\ref{def:ts_to_safety_fw} and let $\counterexample =
\counterexample_0 \ldots, \counterexample_k$ be a counterexample. We
define the game 
\[\tsToGameBw(\ts, \auxHelp, \counterexample, C) = \ltuple C \cup \auxHelp, \tsStates \setminus (C \cup \auxHelp), \gameTransBw, \tsInit, \tsBad \rtuple\] 
\begin{align*} 
\textrm{with} \qquad \counterexample_i \gameTransBw \counterexample_{i+1} & \qquad  \textrm{ for }i \in \{0,\ldots, k {-} 1\}\textrm{ if }\counterexample_i \notin (C \cup \auxHelp) \qquad \textrm{ and} \\ 
s \gameTransBw s' \kern1.1em & \qquad \textrm{ if }s \tsTrans s'\textrm{ and either }s \notin \counterexample\textrm{ or }s \in C \cup \auxHelp. 
\end{align*} 
\end{definition}

The transition relation $\gameTransBw$ in Definition
\ref{def:ts_to_safety_bw} is modified to ``engrave'' the
counterexample. For every state $\counterexample_i$ on the counterexample that is
not part of $C$ or $\auxHelp$, we remove all transitions except for
the one to $\counterexample_{i+1}$. As the counterexample represents a specific
execution, this ensures that we adhere to the behaviour observed
during this execution. By only engraving the counterexample for states
not in $C$ or $\auxHelp$, we selectively analyse which counterexample
states could have changed the outcome by behaving differently.

The safety games defined in Definitions \ref{def:ts_to_safety_fw} and
\ref{def:ts_to_safety_bw} are now used to introduce simple
cooperative games where the players are the actors of the transition
system. In this cooperative game, the value of a coalition $C$ is then
given by the (non-)existence of a winning strategy in the safety game
assigned to $C$.

\begin{definition}[Foward and backward cooperative games] Let $\ts$ be
a transition system and let $\actorSet$ be an actor set. The
\emph{forward cooperative game} is the function $\tsCoopGameFw{\ts}
\colon 2^{\actors} \to \{0, 1\}$ given by \[ \tsCoopGameFw{\ts}(C) =
\gameVal(\tsToGameFw(\ts, \auxHelp, C)). \]

  If $\counterexample$ is a counterexample then the \emph{backward
  cooperative game} is the function
  $\tsCoopGameBw{\ts}{\counterexample} \colon 2^{\actors} \to \{0,
  1\}$ given by \[ \tsCoopGameBw{\ts}{\counterexample}(C) =
  \gameVal(\tsToGameBw(\ts, \auxHelp, \counterexample, C)). \] \end{definition}

The individual contribution of an actor $a$ to the satisfaction 
(and the violation, respectively) of the given safety property can now 
be defined as the Shapley value of $a$ (see Section \ref{sec:prelim} 
for the definition) in the forward and the backward cooperative game,
respectively.

\begin{definition}[Responsibility] \label{def:resp} Let $\ts =
\tsTuple$ be a transition system and $\actorSet$ a responsibility
signature. The \emph{forward responsibility of $\actor \in \actors$}
is $\shapley{\tsCoopGameFw{\ts}}(\actor)$. If $\counterexample$ is a
counterexample then the \emph{backward responsibility of $\actor \in
\actors$} with respect to $\counterexample$ is
$\shapley{\tsCoopGameBw{\ts}{\counterexample}}(\actor)$.
\end{definition}

\begin{remark} The dual of the safety property \emph{``never
$\tsBad$\kern-0.2em''} is the reachability property \emph{``eventually
$\tsBad$\kern-0.2em''}. For forward responsibility, the degree of responsibility
is the same for both the safety and reachability property due to the
zero-sum nature of cooperative games \cite{ForwardResponsibility}. 
For backward responsibility, this is not the case, as the counterexample to 
the safety property does not form a counterexample to the reachability property.
\end{remark}

\begin{remark} The definition of backward responsibility as introduced 
in Definition \ref{def:resp} corresponds to \emph{pessimistic}
responsibility in \cite{BackwardResponsibility} for the case where the
actors are singeltons and constitute a partition of the state space
$S$. Besides pessimistic responsibility, \cite{BackwardResponsibility}
also introduces the concept of \emph{optimistic} responsibility.
Optimistic and pessimistic responsibility differ in how they deal with
the states that are neither on the counterexample nor in $C$. For
pessimistic responsibility, these states are controlled by $\reach$,
whereas for optimistic responsibility, they are controlled by $\safe$.
For optimistic responsibility, \cite{BackwardResponsibility} shows
that for every transition system $\ts$ and counterexample
$\counterexample$, there is a constant $c$ such that every actor
either has responsibility $0$ or $c$. Thus, optimistic responsibility
can be seen as a Boolean concept where each state is either
responsible or not. We focus here on the more interesting (and
computationally harder) concept of pessimistic responsibility.
\end{remark}

\begin{example}
Consider the transition system from Figure~\ref{fig:intro_example}. It
models the tracks leading into a train station. There are two
platforms: $p_1$ and $\lightning$ -- the latter must not be reached,
as it is undergoing maintenance. The switches are controlled by three
actors, called here Alice, Bob and Charlie. Alice controls the
switches $s_0$ and $s_1$, Bob controls $s_2$ and $s_3$ and Charlie
controls $s_4$, $s_5$ and $s_6$. This induces the responsibility
signature $(\{A, B, C\}, \varnothing, \varnothing)$ for $A=\{s_0,
s_1\}$, $B=\{s_2, s_3\}$ and $C=\{s_4, s_5, s_6\}$. We omit $s_7$ and
$\lightning$, as they do not have any outgoing transitions.

	\begin{figure}[htbp]
		\centering
				\begin{tikzpicture}[shorten >=1pt,node distance=1.1cm,on grid,auto, state/.style={circle,inner sep=1pt}]
					\node[state,initial,initial text=] (s0) {$s_0$};
					\node[state] (s1) [above right=of s0]  {$s_1$};
					\node[state] (s2) [below right=of s0]  {$s_2$};
					\node[state] (s3) [below right=of s1]  {$s_3$};
					\node[state] (s4) [above right=of s3]  {$s_4$};
					\node[state] (s5) [right=of s3]  {$s_5$};
					\node[state] (s6) [below right=of s5]  {$s_6$};
					\node[state] (sbad) [above right=of s6]  {$\lightning$};
					\node[state] (s7) [right=of s4]  {$p_1$};
					
					\path[->] (s0) edge (s1);
					\path[->] (s0) edge (s2);
					\path[->] (s1) edge (s3);
					\path[->] (s1) edge (s4);
					\path[->] (s1) edge [bend left] (s7);
					\path[->] (s2) edge (s3);
					\path[->] (s2) edge (s6);
					\path[->] (s3) edge (s4);
					\path[->] (s3) edge (s5);
					\path[->] (s4) edge (s7);
					\path[->] (s4) edge (sbad);
					\path[->] (s5) edge (s6);
					\path[->] (s5) edge (sbad);
					\path[->] (s6) edge (sbad);
		\end{tikzpicture}
	\caption{Model of the tracks leading into a train station. The switches are controlled by the three controllers Alice, Bob and Charlie. The train must not reach $\lightning$, as this track is already occupied.}
	\label{fig:intro_example}
\end{figure}
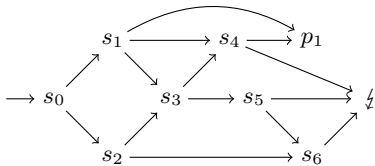

\noindent
With respect to forward responsibility, the pair $(A, \varnothing)$ is
switching: Obviously, $\varnothing$ is not a winning coalition,
whereas $\{A\}$ is by going from $s_0$ to $s_1$ and then to $p_1$,
from where $\lightning$ is unreachable. On the other hand, $(B,
\varnothing)$ and $(C, \varnothing)$ are not switching pairs. Instead,
$(B, \{C\})$ and $(C, \{B\})$ are switching: If $s_2$ is reached, $B$
moves $s_3$, if $s_3$ is reached, $B$ goes to $s_4$ and if $s_4$ is
reached, $C$ goes to $p_1$.

As $B$ and $C$ are not individually winning and $A$ is, we have two
additional switching pairs: $(A, \{B\})$ and $(A, \{C\})$. Given these
switching pairs, computing the Shapley value yields a responsibility
value of $\frac{2}{3}$ for Alice and $\frac{1}{6}$ each for Bob and
Charlie. This makes sense, as Alice is able to route the train to
$p_1$ on her own, whereas Bob and Charlie needs to cooperate to
achieve the same thing, giving them individually less power.
\end{example}

\subsection{Algorithms and complexity} \label{sec:complexity}

The most relevant algorithmic task is the computation of the
responsibility value of a given actor. Besides this task, called  the
\emph{computation problem},
\cite{ForwardResponsibility,BackwardResponsibility} also study the
complexity of the positivity and the threshold problem.

In the \emph{positivity problem}, the question is whether the
responsibility of a given actor is positive. The \emph{threshold
problem} asks whether the responsibility of a given actor is greater
than or equal to a given threshold.

The complexity of these three algorithmic problems has been studied in
\cite{ForwardResponsibility} for the case of forward responsibility
and different types of properties. The results of
\cite{ForwardResponsibility} have then be adapted for the case of
backward responsibility and singleton actors 
\cite{BackwardResponsibility}.
The hardness results presented in \cite{BackwardResponsibility}
carry over to the more general case considered here. The
upper complexity bounds are achieved by similar algorithms as
presented in \cite{ForwardResponsibility} for forward responsibility
and their adaptions presented in \cite{BackwardResponsibility} for
backward responsibility and singleton actors.

As a consequence of
\cite{ForwardResponsibility,BackwardResponsibility}, we obtain the
following results for the setting considered in this article.
For the positivity problem, an NP upper bound is obtained by the
observation that agent $a$ has positive responsibility if and only if
there is at least one switching pair $(C,a)$. Thus, the positivity
problem is solvable by a simple nondeterministic polynomial
time-bounded guess-and-check algorithm that first guesses a coalition
$C \subseteq \Actors \setminus \{a\}$ and then checks whether $(C,a)$
is switching by solving the safety games for the coalitions $C$ and $C
\cup\{a\}$. NP-hardness has been achieved via a reduction from 3SAT.
This yields:

\begin{proposition}[Complexity of the positivity problem]
\label{prop:NP-completeness} The forward positivity problem as well as
the backward positivity problem are NP-complete. \end{proposition}

Given known polynomially time-bounded algorithms for solving safety
games, the threshold and the computation problem are solvable in
exponential time by a na\"ive algorithm that considers all coalitions
$C \subseteq 2^{\Actors}$, computes the winning regions in the induced
safety games and then computes the Shapley values of the agents
according to the definition as a sum that ranges over all coalitions.
This na\"ive approach can be rephrased to obtain polynomially
space-bounded algorithms for the threshold and the computation
problem.

NP-hardness of the threshold problem is a direct consequence of the NP-hardness of the positivity problem (Proposition~\ref{prop:NP-completeness}). \#P-hardness of the computation problem has been shown by reduction from the counting problem of how
many valuations satisfy exactly one clause of a 3SAT formula.

\begin{proposition}[Complexity of the threshold and the computation
problem] \label{prop:threshold_complexity} For both forward and the
backward responsibility, the threshold problem is NP-hard, while the
computation problem is \#P-hard. Furthermore, the threshold and the
computation problem are solvable by polynomially space-bounded
algorithms. \end{proposition}

The NP- and \#P-hardness results are caused by the inherent difficulty
in dealing with the exponential growth of the number of coalitions.
The above sketched algorithm runs exponentially in the number of actors,
but polynomial in the number of states of the transition system. Thus,
the major bottleneck is the size of the actor set rather than the size
of the transition system. If the number of actors is small, then the
computation of responsibility values is feasible even for large state
spaces as our experimental studies (see Section
\ref{sec:implementation}) indicate. In the following section, we present three methods for deriving small actor sets from a system specification. 

%% file: 04-from-reactive-modules.tex
\section{Extracting actors from a reactive system specification}
\label{sec:reactive_modules}

This section presents three methods for deriving actor sets from a program's specification. Transition systems are usually specified in a higher-level modelling language and the methods presented in this section build actor sets from different elements of the modelling language. These elements correspond to semantic units of the final system, so it is natural to group them into actors. Additionally, responsibility values for elements in this higher-level language are easy to interpret, as they abstract the complexity of the underlying transition system.

A popular formalism for program specification are reactive modules~\cite{ReactiveModules}. Here, the program consists of a set of modules, which each have variables and guarded commands. Commands can either be local or synchronised with commands in other modules. Reactive modules make it easy to describe non-determinism, which is why they lend themselves well to modelling transition systems.

In the following, we first introduce the syntax and (informal) semantics of a simple reactive modules language. We then provide schemes for constructing a responsibility based on the modules, variable values and actions of a reactive module program.

\subsection{The reactive modules language}
\label{subsec:reactive_lang}

The syntax of the reactive modules language is given by the following rules:

\begin{bnf*}
	\bnfprod{program}{\lightning \bnfts{=} \bnfpn{bexp} \bnfts{; } \bnfpn{module}^\ast}\\
	\bnfprod{module}{\bnfts{module } \bnfpn{decl}^\ast \bnfpn{command}^\ast \bnfts{ endmodule}  }\\
	\bnfprod{decl}{\bnfpn{id} \bnfts{:[} \bnfpn{int} \bnfts{..} \bnfpn{int} \bnfts{] init } \bnfpn{int} \bnfts{;}} \\
	\bnfprod{command}{\bnfts{[} \bnfpn{act} \bnfts{] } \bnfpn{bexp} \bnfts{->} \bnfpn{updates} \bnfts{;}} \\
	\bnfprod{act}{\varepsilon \bnfor \bnfpn{id}} \\
	\bnfprod{updates}{\varnothing \bnfor (\bnfpn{assignment} \bnfts{ \&})^\ast \bnfpn{assignment}} \\
	\bnfprod{assignment}{\bnfpn{id} \coloneqq \bnfpn{exp}} \\
\end{bnf*}

Here, $\bnfpn{id}$ is an alpha-numeric identifier, $\bnfpn{int}$ is an integer, $\bnfpn{exp}$ is an expression composed of integer and the operators $+$, $-$, $\cdot$, $/$ and $\mathit{mod}$ and $\bnfpn{bexp}$ is a Boolean expression composed of \emph{true}, \emph{false}, the operators $\land$, $\lor$ and $\iff$ applied to Boolean expressions and the operators $=$, $\neq$, $<$, $\leq$, $\geq$ and $>$ applied to expressions.
The left-hand side of an assignment must refer to a variable declared in the same module.

An example is given in Figure~\ref{fig:reactive_modules_language}.
The program has two modules -- \texttt{A} and \texttt{B}. They have variables \texttt{x} and \texttt{y}, respectively. As long as a variable's value is less than $5$, it can be incremented. The command for this does not contain an action, which means it is executed without synchronisation. If there are multiple available commands, one is selected non-deterministically. For example, in state $\{\mathtt{x}\mapsto 3, \mathtt{y} \mapsto 2\}$, there are transitions to $\{\mathtt{x}\mapsto 4, \mathtt{y} \mapsto 2\}$ and $\{\mathtt{x}\mapsto 3, \mathtt{y} \mapsto 3\}$. Additionally, there are commands with synchronising action \texttt{reset}. This can only be executed if both commands are enabled, i.e. if $\mathtt{x}=5$ and $\mathtt{y}=5$. If one module has multiple commands with the same action, one such command with active guard is chosen non-deterministially. A synchronising action may also be used in more than two modules, in which case the action can only be taken if there is an enabled command in each such module. If an action occurs in multiple modules, we call it a synchronising action.

\begin{figure}[ht]
	\centering
	\begin{minipage}[t]{.4\textwidth}
		\centering
		\begin{lstlisting}
module A
	x: [0..5] init 0;
	[] x<5 -> x:=x+1;
	[reset] x=5 -> x:=0;
endmodule
		\end{lstlisting}
	\end{minipage}%
	\begin{minipage}[t]{.4\textwidth}
		\centering
		\begin{lstlisting}
module B
	y: [0..5] init 0;
	[] y<5 -> y:=y+1;
	[reset] y=5 -> y:=0;
endmodule
		\end{lstlisting}
	\end{minipage}
	\caption{Example program in the reactive modules language}
	\label{fig:reactive_modules_language}
\end{figure}

\subsubsection{Semantics}

Let $P$ be a program and $\safetyFormula$ a Boolean expression over the program variables that specifies the invariant of the safety condition ``$\safetyFormula$ is never reached'' (in the following, this is abbreviated to \emph{safety invariant $\safetyFormula$}).
Then the operational semantics of $P$ and $\safetyFormula$ is a transition system $\ts(P, \safetyFormula) = \tsTuple$. We assume that every command in $P$ has an action -- commands with the empty action $\varepsilon$ can be assigned a new, unique action to ensure this holds. The state space $\tsStates$ is the set of variable assignments, i.e. functions that assign to each variable an element of its domain. For $s \in \tsStates$ and variable $v$, we denote by $s(v)$ the value that $v$ takes in $s$. For states $s, s' \in \tsStates$, we have $s \rightarrow s'$ if there exists an action $\alpha$ and a set of commands $C$ such that
\begin{itemize}
	\item every command in $C$ has action $\alpha$,
	\item for every module that contains a command with action $\alpha$, exactly one such command is contained in $C$,
	\item $s$ satisfies the guard of every command in $C$ and
	\item $s'$ is the result of applying the updates of all commands in $C$ to $s$ simultaneously.
\end{itemize}

Finally, $\lightning$ contains those states that satisfy $\safetyFormula$.

\subsection{Module-based actors}
\label{subsec:module_based}

A natural instance of  an actor set of a parallel system is given by its components, i.e., the modules of a reactive module language. 
However, modules specify transitions, not states, so one cannot assign each state to an individual module: In some states, actions from multiple modules may be active. 
Thus, the formalisation of responsibility notions for the modules in our 
state-based approach where actors are sets of states requires some 
transformations.
To this end, we propose adding a scheduler module. Its purpose is to resolve the inter-module nondeterminism, while otherwise preserving the program's structure. As commands with synchronising actions cannot be assigned to a single module, they are handled separately -- the scheduler non-deterministically enables a module or synchronising action.

We need to ensure that the scheduler never chooses a module or synchronising action that is not currently enabled. This would lead to a deadlock that would not be present in the original program. To avoid this, we construct a guard condition for each module and each synchronising action that is only true if the program element can be executed. 

\newcommand{\guard}{\mathsf{Guard}}

\begin{definition}[Module and action guards]
	Let $P$ be a program with modules $M_1, \ldots, M_n$ and synchronising actions $\alpha_1, \ldots, \alpha_m$. Then $\guard(M_i)$ denotes the disjunction of the guards of all commands with non-synchronising action in $M_i$. Furthermore, $\guard(M_i, \alpha_j)$ denotes the disjunction of the guards of all commands with action $\alpha_j$ in $M_i$ and we define $\guard(\alpha_j) = \bigwedge_{i\in \{1, \ldots, n\}}\! \guard(M_i, \alpha_j)$. 
\end{definition}

We now introduce the function $\scheduler$, which takes a program and returns an equivalent program with a scheduler module.

\begin{definition}[Scheduler construction]
	Let $P$ be a program with modules $M_1, \ldots, M_n$ and synchronising actions $\alpha_1, \ldots, \alpha_m$. The scheduler module has a single variable declaration
	
	\quad \texttt{active: [0..n+m] init 0;}\\[0.3ex]
	\noindent For every module $M_i$, the scheduler contains the two commands
	
	\quad \texttt{[choose\_$\mathtt{M_i}$] active=0 \& $\guard(M_i)$ -> active := i;}
	
	\quad \texttt{[act\_$\mathtt{M_i}$] active=i -> active := 0;}\\[0.3ex]
	\noindent For every synchronising action $\mathtt{\alpha_i}$, the scheduler contains the two commands
	
	\quad \texttt{[choose\_$\mathtt{\alpha_i}$] active=0 \& $\guard(\mathtt{\alpha_i})$ -> active := n+i;}
	
	\quad \texttt{[$\mathtt{\alpha_i}$] active=n+i -> active := 0;}\\[0.3ex]
	\noindent The program $\scheduler(P)$ contains all modules from $P$ and the scheduler module. In every module $M_i$, we replace all non-synchronising actions (including the empty action) with $\mathtt{act\_M_i}$.
\end{definition}

The scheduler alternates between two states. When \texttt{active} is $0$, it non-deterministically chooses which module or synchronising action to activate. Note that the guards ensure only modules and synchronising actions are chosen that can then actually be executed. If $1 \leq \mathtt{active} \leq n$, the corresponding module $M_i$ is then activated, because all non-synchronising commands in the module have action \texttt{act\_$M_i$}. Similarly, if $n < \mathtt{active} \leq n + m$, the corresponding synchronising action is executed. This construction preserves safety invariants, as shown in the following lemma.

\begin{lemma}[Soundness of the scheduler transformation]
 \label{lemma:soundness-scheduler}
	Let $P$ be a program and $\safetyFormula$ a safety invariant. Then the semantics of $P$ is given by the transition system $\ts(P, \safetyFormula) = \tsTuple$ and $\lightning$ is reachable in $\ts(P)$ if and only if $\lightning'$ is reachable in $\ts(\scheduler(P), \safetyFormula)=\tsTuple[']$.
\end{lemma}

{
	\newcommand{\transIdx}{\mathcal{I}}
	\newcommand{\rhoAndA}[2]{\rho_{#1}[ a \mapsto #2]}
	\newcommand{\spacer}{\ }
\begin{proof}
	Assume that $\lightning$ is reachable in $\ts(P, \safetyFormula)$. Then there exists some run $\rho = \rho_1 \rho_2 \ldots \rho_n$ with $\rho_n \in \lightning$. For $i \in \{1, \ldots, n - 1\}$, we define $\transIdx(i) = k$ if the transition from $\rho_i$ to $\rho_{i+1}$ was produced by an internal command of module $M_k$ and $\transIdx(i) = n+\ell$ if it was produced by synchronising action $\alpha_\ell$. We abbreviate variable $\mathtt{active}$ with $a$. Then
	\begin{align*}
		\rho' = &\ \rhoAndA{1}{0} \spacer \rhoAndA{1}{\transIdx(1)}\spacer\ldots\spacer\rhoAndA{n-1}{0}\spacer\rhoAndA{n-1}{\transIdx(n-1)}\spacer \rhoAndA{n}{0}
	\end{align*}
	is a run in $\ts(\scheduler(P), \safetyFormula)$ and $ \rhoAndA{n}{0} \in \lightning'$, as $\rho_n \in \lightning$ and thus satisfies $\safetyFormula$. Therefore, $\lightning'$ is reachable in $\ts(\scheduler(P), \safetyFormula)$.
	
	Now assume that $\lightning'$ is reachable in $\ts(\scheduler(P), \safetyFormula)$. The scheduler module alternates between $\mathtt{active}=0$ and $\mathtt{active}>0$. Moreover, as every other module only has synchronising actions with the scheduler module, the scheduler module is active in every step. Therefore, any run in $\ts(\scheduler(P), \safetyFormula)$ has the form $\rho'$ shown above. Then $\rho$ is a run in $\ts(P, \safetyFormula)$ that reaches $\lightning$.
\end{proof}
}

The value of \texttt{active} indicates which module or action is active in a given state. This gives a natural partition of the state-space into module-based actors.

\begin{definition}[Module-based actors]
	Let $P$ be a program with modules $M_1, \ldots, M_n$ and synchronising actions $\alpha_1, \ldots, \alpha_m$. Let $\safetyFormula$ be a safety invariant and let $\tsStates$ be the state space of $\ts(P, \safetyFormula)$. The \emph{module-based responsibility signature} has form $(\actors, \varnothing, \varnothing)$ for $\actors = \{ \actor_i \mid i \in \{0, \ldots, n {+} m\} \}$, where $\actor_i = \{ s \in \tsStates \mid s(\mathtt{active}) = i\}$. We call $\actor_0$ the \emph{scheduler actor}, $\actor_i$ for $1 \leq i \leq n$ the \emph{actor of module $M_i$} and $\actor_{n + j}$ for $1 \leq j \leq m$ the \emph{actor of synchronising $\alpha_j$}.
\end{definition}
%\vspace*{-1\baselineskip}
\begin{figure}[ht]
	\centering
	\begin{minipage}[t]{.5\textwidth}
		\centering
		\begin{lstlisting}
module Window
	// 1: rock-proof glass
	// 2: normal glass
	// 3: broken
	w: [0..3] init 0;
	
	[install] w=0 -> w'=r;
	[a_throws] w=1 -> (§$\varnothing$§);
	[a_throws] w=2 -> w:=3;
	// similar for j_throws
endmodule

module Rebeca
	// 1: rock-proof glass
	// 2: normal glass
	r: [0..2] init 0;
	
	[] r=0 -> r:=1;
	[] r=0 -> r:=2;
	[install] r>0 -> (§$\varnothing$§);
endmodule
		\end{lstlisting}
	\end{minipage}%
	\begin{minipage}[t]{.45\textwidth}
		\centering
		\begin{lstlisting}
module Ada
	// 1: throw!
	// 2: don't throw!
	a: [0..2] init 0; 
	
	[] a=0 -> a:=1;
	[] a=0 -> a:=2;
	[a_throws] a=1 -> a:=2;
endmodule

module Julia
	// 1: throw!
	// 2: don't throw!
	j: [0..2] init 0; 
	
	[] j=0 -> j:=1;
	[] j=0 -> j:=2;
	[j_throws] j=1 -> j:=2;
endmodule
		\end{lstlisting}
	\end{minipage}
	\caption{Model of Ada and Julia throwing rocks at a potentially rock-proof window}
	\label{fig:example_modules}
\end{figure}

\vspace*{-1\baselineskip}
\begin{example}
\label{example:module_based}
	To understand module-based actors, consider the code in Figure~\ref{fig:example_modules}. Here, Ada and Julia may throw a rock at a window. However, Rebeca, the window fitter, may decide to install rock-proof glass. The situation is modelled in such a way that these choices are first made locally in the module and then transferred to the window after that using synchronising actions.
	
	In our scenario, Rebeca first installs normal (non-rock-proof) glass, after which Ada decides that she wants to throw a rock, breaking the window. Analysing responsibility yields the following for the module actors: $\mathit{Ada} \mapsto \frac{1}{6}$,  $\mathit{Julia} \mapsto \frac{1}{6}$ and  $\mathit{Rebeca} \mapsto \frac{2}{3}$. This is because Rebeca could have prevented the window from being broken on her own. Meanwhile, even if Ada decided not to throw her rock, Julia might still throw her rock, so they need to collaborate to prevent the window from breaking.
	In this example, neither the scheduler nor the synchronising actions have any responsibility.
\end{example}

\subsection{Value-based actors}
\label{subsec:value_based}

We now show how actors can be constructed based on the values of some of the program variables. For this, we select some variables and then place those states in the actor where the chosen variables have the same value. 

\begin{definition}[Value-based actors]
	Let $P$ be a program, $\safetyFormula$ a 
safety invariant, $V = \{v_1, \ldots, v_k\}$ some of the variables of $P$ with domains $D_1, \ldots,D_k$ and $\tsStates$ the state space of $\ts(P, \safetyFormula)$. The \emph{value-based responsibility signature} has form $(\actors, \varnothing, \varnothing)$ for 
\[
  \actors \ = \ 
  \bigl\{ \, a_{(d_1,\ldots,d_k)} \, \mid \, (d_1,\ldots,d_k) \in D_1 \times \ldots \times D_k \, \bigr\}
\]
where
$a_{(d_1, \ldots, d_k)} = \bigl\{ s \in \tsStates \mid s(v_i) = d_i, i=1, \ldots, k\bigr\}$.
\end{definition}

\begin{remark}
	Module-based actor groups are a special case of value-based actors, where $V =\{ \mathtt{active}\}$ for the variable \texttt{active} of the scheduler.
\end{remark}

One obvious instance of value-based actor sets is obtained taking V to be the set of all control variables (program counters) or a subset thereof. The induced responsibility notions can help to identify the control locations that are most critical for the satisfaction of the safety property.

\begin{example}
	\begin{figure}
		\centering
		\begin{minipage}{.44\textwidth}
			\begin{lstlisting}[extendedchars=true,literate={ö}{{\"o}}1]
module Clock
	t: [8..20] init 8;
	
	[tick] t<20 -> t:=t+1;
endmodule

module Car
	loc: [0..13] init 0;
	
	// Malmö
	[tick] loc=0 -> loc=1;
	[tick] loc=0 -> loc=2;
	
	// Helsingborg
	[tick] loc=1 -> loc=3;
	[tick] loc=1 -> loc=4;
	
	// ... (see right)
endmodule\end{lstlisting}
		\end{minipage}%
		\begin{minipage}{.55\textwidth}
			\centering
			\begin{tikzpicture}[shorten >=1pt,auto, state/.style={rectangle,inner sep=1pt},x=1.0cm,y=1.0cm]
				\node[state,initial,initial text=]  at (0,0.2) (Malmo) {Malmö};
				\node[state]  at (-0.6,1) (Helsingborg) {Helsingborg};
				\node[state]  at (-1.8,2.8) (Goteborg) {Göteborg};
				\node[state]  at (-2.7,6.0) (Oslo) {Oslo};
				\node[state]  at (-1.5,5.3) (Karlstad) {Karlstad};
				\node[state]  at (0.2,5.1) (Orebro) {Örebro}; 
				\node[state]  at (-0.1,2.9) (Jonkoping) {Jönköping};
				\node[state]  at (1.4, 0.1) (Ystad) {Ystad};
				\node[state]  at (0.8, 3.5) (Linkoping) {Linköping};
				\node[state]  at (1.6, 4.1) (Norrkoping) {Norrköping};
				\node[state]  at (2.8, 4.7) (Stockholm) {Stockholm};
				\node[state]  at (1.7, 5.4) (Vasteras) {Västerås};
				\node[state]  at (3.0, 7.3) (Lulea) {Luleå};
			
				\path[->] (Malmo) edge (Ystad);
				\path[->] (Malmo) edge (Helsingborg);
				\path[->] (Malmo) edge [bend right=30] (Norrkoping);
				\path[->] (Helsingborg) edge (Goteborg);
				\path[->] (Helsingborg) edge (Jonkoping);
				\path[->] (Goteborg) edge (Oslo);
				\path[->] (Goteborg) edge (Karlstad);
				\path[->] (Goteborg) edge (Orebro);
				\path[->] (Orebro) edge (Vasteras);
				\path[->] (Jonkoping) edge (Linkoping);
				\path[->] (Linkoping) edge (Norrkoping);
				\path[->] (Norrkoping) edge (Stockholm);
				\path[->] (Stockholm) edge (Lulea);
				\path[->] (Stockholm) edge (Vasteras);
				\path[->] (Jonkoping) edge [bend left=10] (Orebro);
				\path[->] (Karlstad) edge (Orebro);
				\path[->] (Ystad) edge [loop right] (Ystad);
				\path[->] (Oslo) edge [loop above, looseness=23] (Oslo);
				\path[->] (Lulea) edge [loop left] (Lulea);
				\path[->] (Vasteras) edge [loop above, looseness=23] (Vasteras);
			\end{tikzpicture}
		\end{minipage}
		\caption{Model of Ada and Rebeca's drive from Malmö to Västerås.}
		\label{fig:example_values}
	\end{figure}
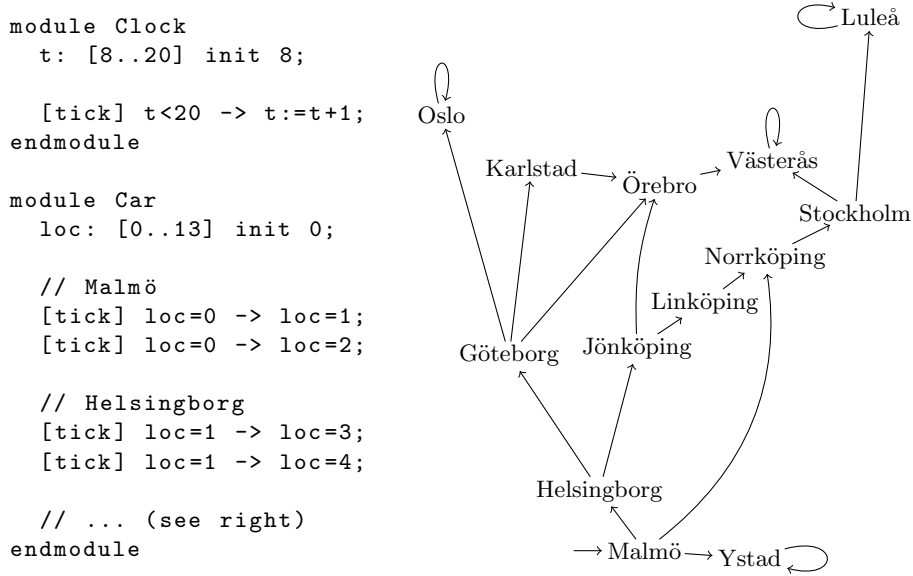
	Ada and Rebeca are driving from Malmö to Västerås. Ada is behind the wheel, but does not know Swedish geography, while Rebeca knows the way, but is very tired. Rebeca therefore wants to know when she needs to be awake. For this, she models the major roads in southern Sweden. The code for this is given in Figure~\ref{fig:example_values}, as is a graph that shows the area modelled by Rebeca. 
	
	 They start at 8:00 and want to arrive by 20:00. Rebeca therefore uses the property ``never $\texttt{t}=20 \land \texttt{loc}\neq\texttt{Västerås}$'', where variable \texttt{t} tracks the current time and \texttt{loc} tracks the current location. She groups states based on the value of \texttt{time} and computes the forward responsibility.
	
	{
		\newcommand{\tAct}[1]{\texttt{time}=#1}
		This reveals that the actors $\tAct{8}$, $\tAct{9}$, $\tAct{10}$ and $\tAct{13}$ have positive responsibility. Therefore, Rebeca can safely sleep in the hours between that, as it is impossible for any important decisions to occur in these times. The actor $\tAct{13}$ has positive responsibility because the route Malmö $\to$ Helsingborg $\to$ Jönköping $\to$ Linköping $\to$ Norrköping $\to$ Stockholm reaches Stockholm at 13:00. Rebeca needs to ensure that Ada takes the correct turn in Stockholm to avoid travelling to Luleå in Sweden's arctic north.
	}
\end{example}

\subsection{Action-based actors}
\label{subsec:action_based}

Module-based responsibility is often not sufficiently granular and value-based responsibility only makes sense in models where a few distinguished variables represent different components of a program. In this section, we propose a method for generating actors that correspond to the actions in a program. However, there is no one-to-one correspondence between actions in a program and the states in its semantics. Furthermore, unlike for modules, we cannot apply the scheduler construction like we did for modules: This would transfer all non-deterministic choices of the program to the scheduler, thus providing no insight into which actions are responsible for an outcome.

We propose a transformation on transition systems that makes the responsibility of each action explicit. We then define a responsibility signature where each actor corresponds to an action.

\subsubsection{The action-separation construction.}

Most states in a transition system have multiple outgoing transitions corresponding to different actions. Our goal is to determine which of these actions are responsible for whether the property is fulfilled. For this, we separate the actions of each state into separate auxiliary states. This allows us to individually measure whether each action has an impact on the satisfaction of the property.

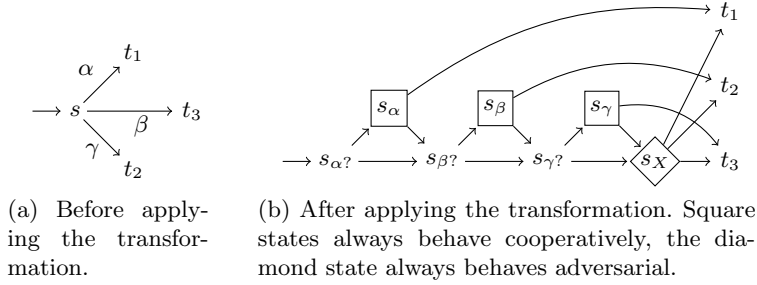
\begin{figure}
	\centering
	\subfloat[Before applying the transformation.]{{
			\label{fig:action_separation_example:before}
			\begin{tikzpicture}[shorten >=1pt,node distance=1.1cm,on grid,auto, state/.style={circle,inner sep=1pt}] 
				\node[state,initial,initial text=] (s)   {$s$}; 
				\node[state] (t1) [above right=of s] {$t_1$}; 
				\node[state] (t2) [below right=of s] {$t_2$}; 
				\node[state](t3) [below right=of t1] {$t_3$};
				\path[->] (s) edge node  {$\alpha$} (t1);
				\path[->] (s) edge node[below right=-0.42cm and 0.3cm]  {$\beta$} (t2);
				\path[->] (s) edge node[below left=0.32cm and 0.3cm] {$\gamma$} (t3);
	\end{tikzpicture}}}%
	\qquad
	\subfloat[After applying the transformation. Square states always behave cooperatively, the diamond state always behaves adversarial.]{{
			\label{fig:action_separation_example:after}
			\begin{tikzpicture}[shorten >=1pt,node distance=1.0cm,on grid,auto, state/.style={circle,inner sep=1pt},friendly/.style={draw, regular polygon,regular polygon sides=4, inner sep=0}, opp/.style={draw, regular polygon, regular polygon sides=4, shape border rotate=45, inner sep=-1pt}]
				\node[state,initial,initial text=] (saq)   {$s_{\alpha?}$};
				\node[state, friendly] (sa) [above right=of saq] {$s_{\alpha}$}; 
				
				\node[state] (sbq) [below right=of sa] {$s_{\beta?}$};
				\node[state, friendly] (sb) [above right=of sbq] {$s_{\beta}$};
				
				\node[state] (scq) [below right=of sb] {$s_{\gamma?}$};
				\node[state, friendly] (sc) [above right=of scq] {$s_{\gamma}$};
				
				\node[state, opp] (sx) [below right=of sc] {$s_{X}$};
				
				\node[state] (t1) [above right=2cm and 1.0cm of sx] {$t_1$};
				\node[state] (t2) [below=of t1] {$t_2$};
				\node[state] (t3) [below=of t2] {$t_3$};
				
				\path[->] (saq) edge (sa);
				\path[->] (saq) edge (sbq);
				\path[->] (sa) edge (sbq);
				
				\path[->] (sbq) edge (sb);
				\path[->] (sbq) edge (scq);
				\path[->] (sb) edge (scq);
				
				\path[->] (scq) edge (sc);
				\path[->] (scq) edge (sx);
				\path[->] (sc) edge (sx);
				
				\path[->] (sa) edge [bend left=20] (t1);
				\path[->] (sb) edge [bend left=25] (t2);
				\path[->] (sc) edge [bend left] (t3);
				
				\path[->] (sx) edge (t1);
				\path[->] (sx) edge (t2);
				\path[->] (sx) edge (t3);
				
			\end{tikzpicture}
	}}%
	\caption{Action-separation transformation applied to a state with three actions.}%
	\label{fig:action_separation_example}%
\end{figure}

We first give an intuition of the technique and then provide a formal definition. The construction is illustrated in Figure~\ref{fig:action_separation_example} for a single state $s$ in a transition system. Figure~\ref{fig:action_separation_example:before} shows the state in the original transition system. It has three outgoing transitions $\alpha$, $\beta$ and $\gamma$ to three new states $t_1$, $t_2$ and $t_3$.  The transformation is depicted in Figure~\ref{fig:action_separation_example:after}. State $s$ has been expanded into several states. The states $s_{\alpha?}$ provides two options: Either we enable action $\alpha$ and transition to $s_{\alpha}$ or we directly proceed to $s_{\beta?}$, which provides the same two options for $\beta$. If we choose to enable action $\alpha$, we can choose action $\alpha$ in state $s_{\alpha}$, which leads to state $t_1$ as in the original transition system. However, we can also choose to not take $\alpha$ and proceed in $s_{\beta?}$. These states are depicted as squares because they always behave cooperatively, i.e., they always try to fulfil the safety property, regardless of whether they are in the coalition.

If we have not taken any action, we finally reach to $s_{X}$. This state is depicted as a diamond to symbolise that it always behaves adversarially, i.e., it always tries to violate the safety invariant.

Because states $s_{\alpha}$, $s_{\beta}$, $s_{\gamma}$ and $s_{X}$ have fixed behaviour, they never have positive responsibility. State $s_{\alpha?}$ has positive responsibility if enabling $\alpha$ is critical for satisfying the property in some context $C$. For this, it is necessary that the context $C$ and $\alpha$ together are sufficient for enforcing the property and that the context alone is insufficient.

Note that the second condition does not hold if both $\beta$ and $\gamma$ are also sufficient for enforcing the property in context $C$: The demonic state $s_{X}$ must choose one of the actions and if every choice leads to satisfying the property, then none of the actions are responsible. This also means that an action in a state with a single action never has any responsibility.

Having introduced the intuition for a single state, we now provide a formalisation for an entire transition system.

{
	\newcommand{\firstAction}{\mathsf{First}}
	\newcommand{\nextAction}{\mathsf{Next}}
	\begin{definition}[Action-separation construction]
		Let $P$ be a program and $\safetyFormula$ a safety invariant. Then the semantics of $P$ is $\ts(P, \safetyFormula) = \tsTuple$. Assume that $P$ has no unnamed actions. Let $\ell$ be a function that maps each transition to the action that produced it. If there are multiple actions that produce the same transition, we arbitrarily choose one of them. The set of actions available in $s$ is denoted by $\acts(s) = \{ \alpha \in \acts \mid \exists t \in \tsStates \colon s \tsTrans t, \ell((s, t)) = \alpha \}$.
		
		The function $\actionSeparated(\ts) = \tsTuple[']$ produces a transition system where the actions are separated. Here, $\tsStates' = \tsStates_? \cup \tsStates_! \cup \tsStates_X$, where $\tsStates_? = \{ s_{\alpha?} \mid s \in \tsStates, \alpha \in \acts(s)  \}$,  $\tsStates_! = \{ s_{\alpha} \mid s \in \tsStates, \alpha \in \acts(s)  \}$ and  $\tsStates_X = \{ s_{X} \mid s \in \tsStates \}$. Given an arbitrary ordering of the actions $\alpha_1, \ldots, \alpha_k$ of $s \in \tsStates$, we denote by $\firstAction(s) = s_{\alpha_1?}$ the first ?-state and by $\nextAction$ with $\nextAction(s, \alpha_i) = s_{\alpha_{i+1}?}$ for $i < k$ and $\nextAction(s, \alpha_k) = s_{X}$ the ?-state of the next action (or $s_X$ if the last action is reached).
		
		\noindent
		The transition relation $\tsTrans'$ is composed of the following elements for each $s \in \tsStates$:
		\begin{itemize}
			\item $s_{\alpha?} \tsTrans s_{\alpha}$ for $\alpha \in \acts(s)$,
			\item $s_{\alpha?} \tsTrans \nextAction(s, \alpha)$ for $\alpha \in \acts(s)$,
			\item $s_{\alpha} \tsTrans t$ for $(s, t) \in \tsTrans, \ell((s, t)) = \alpha$,
			\item $s_{\alpha} \tsTrans \nextAction(s, \alpha)$ for $\alpha \in \acts(s)$ and
			\item $s_{X} \tsTrans t$ for $(s, t) \in \tsTrans$.
		\end{itemize}
	
	Furthermore, $\tsInit' = \firstAction(\tsInit)$ and $\tsBad' = \{ \firstAction(s) \mid s \in \tsBad\}$
	\end{definition}

Similar to the scheduler transformation for module-based actors, one can show that this transformation preserves safety invariants.

\begin{lemma}%[Preservation of safety properties]
   [Soundness of the action-separation transformation]
  \label{lemma:soundness-action}
	Let $P$ and $\safetyFormula$ be as above with semantics $\ts(P, \safetyFormula) = \tsTuple$. The action-separation transformation yields transition system $\actionSeparated(\ts(P, \safetyFormula)) = \tsTuple[']$. Then $\tsBad$ is reachable in $\ts(P, \safetyFormula)$ if and only if $\tsBad'$ is reachable in $\actionSeparated(\ts(P, \safetyFormula))$.
\end{lemma}

\newcommand{\rhoLast}{\rho'_{\mathit{last}}}
\begin{proof}
	Assume that $\tsBad$ is reachable in $\ts(P, \safetyFormula)$. Then there exists a loop-free run $\rho = \rho_1 \ldots \rho_n$ with $\rho_n \in \tsBad$. For $i \in \{1, \ldots, n - 1\}$, we define $\act(i) = \alpha$ when $\ell((\rho_i, \rho_{i+1})) = \alpha$. The run $\rho'$ that reaches $\tsBad'$ in $\actionSeparated(\ts(P, \safetyFormula))$ is constructed as follows. Here, $\rhoLast$ refers to the last state of $\rho'$.
	
	For every $i \in \{1, \ldots, n - 1\}$, we first append $\firstAction(\rho_i)$. This is a state of form $\rho_{i, \alpha?}$ for some action $\alpha$. While $\rhoLast = \rho_{i, \alpha?}$ for $\alpha \neq \act(i)$, we append $\nextAction(\rhoLast, \alpha)$. Eventually, we have $\alpha=\act(i)$, as $\alpha$ is an action in state $\rho_i$. We then append $\rho_{i, \alpha}$. By construction, there exists a transition to $\firstAction(\rho_{i+1})$. Finally, we append $\firstAction(\rho_n)$ and have $\rhoLast \in \tsBad'$ in $\actionSeparated(\ts(P, \safetyFormula))$ because $\rho_n \in \tsBad$ in $\ts(P, \safetyFormula)$.
	
	Now assume that $\tsBad'$ is reachable in $\actionSeparated(\ts(P, \safetyFormula))$. Then there exists a loop-free run $\rho = \rho_1 \ldots \rho_n$ with $\rho_n \in \tsBad'$. We construct a run $\rho'$ that reaches $\tsBad$ in $\ts(P, \safetyFormula)$ as follows. We sequentially consider every state $\rho_i \in \rho$. If there is an $s \in \tsStates$ such that $\rho_i = \firstAction(s)$, we append $s$ to $\rho'$. Otherwise, we proceed without appending any state to $\rho'$.
	
	As the last state $\rho_{\mathit{last}}$ of $\rho$ reaches $\tsBad'$, it has form $\rho_{\mathit{last}} = \firstAction(s)$ for some $s \in \tsStates$ with $s \in \lightning$. Therefore, $s$ is the last state of $\rho'$ and $\rho'$ therefore reaches $\tsBad$ in $\ts(P, \safetyFormula)$.
	
	It remains to show that $\rho'$ is a run $\ts(P, \safetyFormula)$. The initial state of the transition system $\actionSeparated(\ts(P, \safetyFormula))$ is $\firstAction(\tsInit)$, so $\tsInit$ is the first state in $\rho'$. We now show that $\rho'$ adheres to the transition relation $\tsTrans$.
	
	Consider consecutive states $s, t \in \rho'$. We show that $(s, t)\in \tsTrans$. Let $\rho_i, \rho_j \in \rho$ be the states with $\firstAction(s) = \rho_i$ and $\firstAction(t) = \rho_j$. Note that $\rho$ is loop-free and therefore, $\rho_i$ and $\rho_j$ are well-defined.
	
	From $\rho_i$, one can only reach states of the forms $s_{\alpha?}$, $s_{\alpha}$ and $s_X$ (for action $\alpha$) and states of form $\firstAction(t')$ for $t' \in \tsStates$. Therefore, all states between $\rho_i$ and $\rho_j$ have the form $s_{\alpha?}$, $s_{\alpha}$ or $s_X$. Now consider the state $\rho_{j-1}$. We have $\rho_{j-1} \neq s_{\alpha?}$ for any action $\alpha$, as ?-states do not have transitions to states of form $\firstAction(t')$ for any $t' \in \tsStates$. If we have $\rho_{j-1} = s_X$, then there is a transition between $s_X$ and $\rho_{j}$ in $\actionSeparated(\ts(P, \safetyFormula))$. This implies that there is a transition between $s$ and $t$  in $\ts(P, \safetyFormula)$. Alternatively, $\rho_{j-1}=s_{\alpha}$. In this case, there is a transition between $s_{\alpha}$ and $\rho_j$ in $\actionSeparated(\ts(P, \safetyFormula))$. This once again implies that there is a transition between $s$ and $t$ in $\ts(P, \safetyFormula)$. Therefore, $\rho$ is a run and we have proved the lemma.
\end{proof}
}

\begin{definition}[Action-based actors]
	Let $P$ be a program with actions $\acts$ and let $\safetyFormula$ be a safety invariant. Let $\ts(P, \safetyFormula) = \tsTuple$ be the semantics of $P$. The \emph{action-based responsibility signature} has form $(\actors, \auxHelp, \auxAdv)$ for
	\begin{align*}
		\actors\ =\ & \{ a_\alpha \mid \alpha \in \acts \}  \textrm{ with } a_\alpha = \{ s_{\alpha?} \mid s \in \tsStates \}, \\
		\auxHelp\ =\ & \{ s_{\alpha} \mid \alpha \in \acts, s \in \tsStates \}, \\
		\auxAdv\ =\ & \{ s_{X} \mid s \in \tsStates \}.
	\end{align*}
\end{definition}

\begin{example}
	For her birthday, Rebeca was gifted a puzzle box. The box displays a number -- initially $0$ -- and has three buttons: The first button increments the number by $2$, the second one multiplies the number by $6$ and then increments it by $1$ and the third button squares the current number. The box will open and reveal Rebeca's gift if, after exactly twenty button presses, the display shows $60$. Rebeca is wondering whether all the buttons are useful for reaching this goal.
	
	\begin{figure}
		\centering
		\begin{lstlisting}
module PuzzleBox
	counter: [0..61] init 0;
	steps: [0..20] init 0;
	
	[btn1] true -> counter:=counter+2 & steps:=steps+1;
	[btn2] true -> counter:=counter*7+1 & steps:=steps+1;
	[btn3] true -> counter:=counter*counter & steps:=steps+1;
endmodule
		\end{lstlisting}
		\caption{Model of Rebeca's puzzle box. For the sake of simplicity, overflow checks are omitted here.}
		\label{fig:example_actions}
	\end{figure}
	
	To find out, she models the entire box as a reactive module. The code is shown in Figure~\ref{fig:example_actions}. We assume that values larger than 61 are clamped to 60 in the model (this is omitted for simplicity in the code shown). This does not affect the model's semantics, as none of the actions decrease the value of \texttt{counter}. Therefore, 60 is unreachable as soon as a value larger than 60 is reached.
	
	Rebeca chooses the safety property ``\emph{never} $\texttt{counter}\neq60 \land \texttt{steps}=20$'' and computes forward action-based responsibility for the three actions \texttt{btn1}, \texttt{btn2}, \texttt{btn3}. This reveals that \texttt{btn2} has no responsibility, whereas the other two actions both have responsibility $\frac{1}{2}$.
	
	This result is to be expected, since the goal is to reach an even number. Pressing the third button always produces an odd number and none of the buttons give a way to turn an odd number into an even number. Therefore, pressing the fourth button is never necessary when trying to reach $60$. Armed with this knowledge, Rebeca is quickly able to solve the box and get to her gift.
\end{example}

%% file: 06-experiments.tex
\section{Experimental evaluation}
\label{sec:implementation}

In this section, we report on initial experiments for actor-based responsibility. 
We use the prototypical implementation from \cite{BackwardResponsibility} enriched
with functionality to support actors. Our tool relies on the probabilistic model checker 
\textsc{Prism}~\cite{Prism} for building the transition system and for counterexample 
generation. The input language of \textsc{Prism} is similar to the guarded command 
language introduced in Section~\ref{sec:reactive_modules}. It is possible to specify stochastic behaviour in \textsc{Prism}'s language, but we do not make use of this. Actors can be specified manually 
or extracted from the \textsc{Prism} source code automatically using the techniques 
described in Section~\ref{sec:reactive_modules}. The implementation supports the 
na\"ive approach of iterating over all coalitions to compute the responsibility values. 
In addition to this, the tool contains a randomised procedure, which samples coalitions 
and approximate the responsibility values rather than computing the precise values.

The experimental studies as reported in this section investigate how responsibility 
is distributed among different actors when considering the three different types of 
actor classes. We study the general applicability of the approach and show how 
responsibility values can be used to guide debugging and diagonstic processes. Furthermore, 
we investigate the scalability of our approach for an increasing number reachable states 
in the transition system and increasing numbers of actors.
All experiments were performed on a MacBook Pro running macOS 13.3 with an 8-core M2 
chip and 24 GB of memory. The code, example files (small examples from the previous 
sections), as well as all relevant data files are made available for download via \cite{Implementation}. 

\subsection{Case study: bounded retransmission protocol}
The first case study is on a non-probabilistic (i.e. purely a non-deterministic) variant
of the bounded retransmission protocol \cite{PrismBenchmarkSuite} from the 
\textsc{Prism} benchmark suite. The goal is to send a message, split into a number 
of frames, via an unreliable communication channel. The \textsc{Prism} code contains 
modules for a sender, a receiver, two communication channels and a checker module, 
which determines whether the message was transmitted correctly. If transmission fails, 
the frame is to be re-transmitted, but only a bounded number of times. When the number 
of retransmissions reaches its maximum, the protocol aborts. Both sender and receiver 
can be different locations, for example \texttt{idle}, \texttt{next\_frame} or \texttt{retransmit}. 
Those locations are encoded as values of respective variables. Those values are 
used to define our variable-value-based actors. Messages can be lost non-deterministically at different locations during the
execution of the protocol. 

We consider here a counterexample in which sending fails
repeatedly during the transmission of the first frame, causing the message transmission to ultimately fail when the maximum number of retries is reached. Our analysis reveals that the location \texttt{wait\_ack} 
has responsibility $\frac{2}{3}$, while \texttt{next\_frame} and \texttt{retransmit} both 
have responsibility $\frac{1}{6}$. This is unexpected, as the sender module does not 
behave non-deterministically in the \texttt{wait\_ack} location. In a second investigation, 
we therefore construct actors based on the location of the sender and the transmission channel. 
This reveals that the responsible actors all represent the beginning of transmission in the 
channel. This hints at some problem in the channel -- and indeed, the channel may 
non-deterministically lose the frame, as it is the case in our counterexample.

\subsection{Analysis of the scheduler construction}

Module-based actors divide responsibility between a scheduler module, synchronising 
actions and the modules of the program. In Example~\ref{example:module_based}
only the modules have positive responsibility, whereas scheduler and shared actions have 
none. In this section, we evaluate how the scheduler construction allocates responsibility 
for examples taken from \cite{BackwardResponsibility} and in parts originating in the 
QComp benchmark suite \cite{QComp}, again adapted to the non-probabilistic setting.

\begin{figure}
	\begin{center}
		\setlength{\tabcolsep}{6pt}
		\begin{tabular}{ r r r r }
			\textbf{Name} & \makecell{\textbf{Scheduler}\\\textbf{module}} & \makecell{\textbf{Synchronous}\\\textbf{actions}} & \textbf{Modules} \\ 
			\hline
			\texttt{alternating\_bit} & 0.50  & 0.00 & 0.50 \\ 
			\texttt{brp} & 0.00  & 1.00 & 0.00 \\
			\hline
			\texttt{dining\_philosophers\_a} & 1.00 & 0.00 & 0.00 \\
			\texttt{dining\_philosophers\_b} & 0.50 & 0.00 &0.50 \\
			\hline
			\texttt{3\_generals\_a} & 1.00 & 0.00 & 0.00 \\
			\texttt{3\_generals\_b} & 0.08 & 0.00 & 0.92 %Actual values: 0.083333... and 0.91666...
		\end{tabular}
	\end{center}
	\caption{Distribution of module-based responsibility among the scheduler module, 
	the synchronising actions and the modules of the respective model.
	}
	\label{fig:module_based_distribution_table}
\end{figure}

The results are shown in Figure~\ref{fig:module_based_distribution_table}. 
The model \texttt{brp} is the only one where synchronising actions have positive 
responsibility. This is because \texttt{brp} is the only model among our examples where multiple commands with the same synchronising action may be enabled simultaneously. In the other modules, this is not the case. Therefore, once the scheduler has chosen a synchronising action, the next step is deterministic -- only one command (in every module) is enabled.

For \texttt{dining\_philosophers} and \texttt{3\_generals}, we provide two variants. 
The original versions with suffix \texttt{\_a} allocate all responsibility to the scheduler. 
To understand why, consider one philosopher in \texttt{dining\_philosopher\_a}. 
Assuming she does not hold any forks, she has two available commands:
\nopagebreak

\qquad \texttt{[Phil1TakesFork1] !has\_fork\_1 -> has\_fork\_1 := true;}

\nopagebreak
\qquad \texttt{[Phil1TakesFork2] !has\_fork\_2 -> has\_fork\_2 := true;}\\[0.3ex]
The forks are modelled by additional modules that provide commands with 
actions \texttt{Phil1TakesFork1} and \texttt{Phil1TakesFork2}, making them 
synchronising actions. Therefore, by choosing one of these actions, the scheduler 
has already resolved all non-determinism and the behavior of the philosopher and the 
fork are fully deterministic.
In \texttt{dining\_philosophers\_b}, on the other hand, the fork selection process 
is split into two stages. In the first stage, the philosopher chooses which fork 
she wants and stores the result in an internal variable. In a second step, the 
philosopher then uses a synchronising action to transmit that she took the fork. 
Now, the scheduler only chooses which philosopher acts. The choice between left and 
right fork is made in the philosopher module. Therefore, it makes sense that 
responsibility is split between the scheduler and the modules here.

For \texttt{3\_generals}, the explanation is similar. In the original model, choices 
are made and transmission happens simultaneously, whereas in the modified protocol, 
choices are made locally first and then transmission happens.

These examples demonstrate that module-based responsibility is very sensitive 
to the specific modelling details. If only the scheduler is responsible, it is often 
possible to refine the model to better represent where choices are actually made.

\subsection{Performance}

As outlined in Section~\ref{sec:complexity}, computing responsibility is 
exponential in the number of actors and linear in the size of the transition system. 
In the following, we investigate how the tool performs in practice. For this, we 
investigate the time taken to actually \emph{compute} responsibility only\footnote{
Note that we do not report here on the overhead runtimes for exporting and re-importing the 
model. %Parsing the input currently dominates the computation time (in some cases, 
%it is more than 140 times slower than the computation), 
The responsibility computation is not (yet) fully integrated into \textsc{prism} 
or any other model checker, which makes the current prototypical implementation rather 
inefficient at this point. An integration would avoid the expensive and unnecessary 
data exchange via the file system.}.

We use the following three model families in this investigation. For all three,
the states emerge from the evaluation of a variable ranging from $0$ to $n$.
\begin{itemize}
	\item \texttt{linear}: each state has several transitions which increment the 
	variable value by different step sizes. The goal is to avoid the state where the variable 
	value is $n$. The state space is partitioned into $m$ actors such that states 
	with the same index modulo $m$ belong to he same actor.
	\item \texttt{random}: every state has transitions to six random other states. 
	Several states are marked as $\lightning$ and we add a few additional transitions 
	to ensure a counterexample exists. The states are randomly partitioned into 
	evenly-size actors.
	\item \texttt{tree}: states are arranged as a binary tree. The root is the initial 
	state, every tenth leaf is in $\lightning$ and states are partitioned into evenly-sized 
	actors.
\end{itemize}

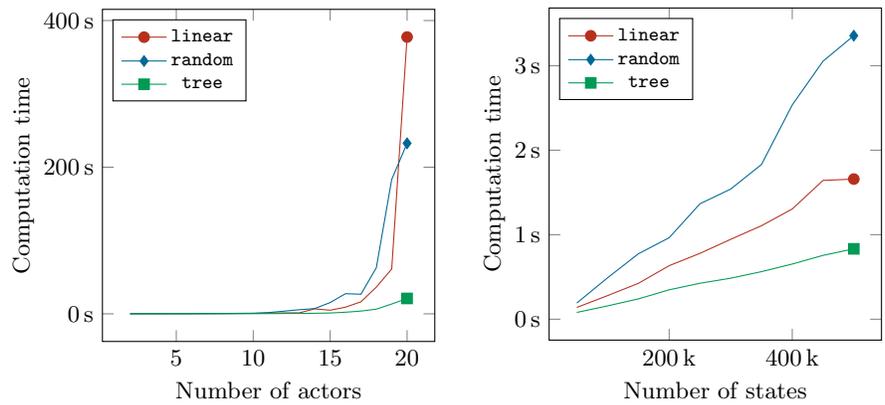
\begin{figure}
	\centering
	\subfloat[Runtime for different number of actors and constant state space size 100~k.]{
		\label{fig:performance_actors}
		\begin{tikzpicture}
			\begin{axis}[
				xlabel=Number of actors,
				ylabel=Computation time,
				ylabel shift=-0.1cm,
				xticklabel = \pgfmathprintnumber{\tick},
				yticklabel = \pgfmathprintnumber{\tick}\,s,
				width=6cm,height=6cm,
				legend style={nodes={scale=0.83, transform shape}},
				legend pos=north west,
				legend image post style={mark indices={}},
				]
				\addplot[BrickRed, mark=*,mark indices=19] plot coordinates {
					(2, 0.098)
					(3, 0.123)
					(4, 0.16)
					(5, 0.193)
					(6, 0.223)
					(7, 0.28)
					(8, 0.35)
					(9, 0.448)
					(10, 0.655)
					(11, 0.918)
					(12, 1.323)
					(13, 1.373)
					(14, 6.815)
					(15, 4.898)
					(16, 9.085)
					(17, 16.42)
					(18, 36.41)
					(19, 61.165)
					(20, 377.503)
				};
			\addlegendentry{\texttt{linear}}
				\addplot[MidnightBlue, mark=diamond*,mark indices=19] plot coordinates {
					(2,0.135)
					(3,0.183)
					(4,0.23)
					(5,0.278)
					(6,0.365)
					(7,0.433)
					(8,0.535)
					(9,0.69)
					(10,0.815)
					(11,1.76)
					(12,3.65)
					(13,5.63)
					(14,7.18)
					(15,15.4)
					(16,27.49)
					(17,26.72)
					(18,62.86)
					(19,183.3)
					(20,232.6)
				};
			\addlegendentry{\texttt{random}}
				\addplot[ForestGreen, mark=square*,mark indices=19] plot coordinates {
					(2, 0.075)
					(3, 0.0775)
					(4, 0.1)
					(5, 0.12)
					(6, 0.14)
					(7, 0.1625)
					(8, 0.19)
					(9, 0.2225)
					(10, 0.2475)
					(11, 0.3525)
					(12, 0.4925)
					(13, 0.545)
					(14, 0.7525)
					(15, 1.16)
					(16, 2.21)
					(17, 3.765)
					(18, 6.3975)
					(19, 13.4075)
					(20, 21.005)
				};
				\addlegendentry{\texttt{tree}}
			\end{axis}
		\end{tikzpicture}
	}
\quad
	\subfloat[Runtime for different state space size and fixed number of 7~actors.]{
		\label{fig:performance_states}
		\begin{tikzpicture}
				\begin{axis}[
				xlabel=Number of states,
				ylabel=Computation time,
				ylabel shift=-0.1cm,
				xticklabel = \pgfmathprintnumber{\tick}\,k,
				yticklabel = \pgfmathprintnumber{\tick}\,s,
				width=6cm,height=6cm,
				legend style={nodes={scale=0.83, transform shape}},
				legend pos=north west,
				legend image post style={mark indices={}}
				]
				\addplot[BrickRed, mark=*,mark indices=10] plot coordinates {
					(50,0.139)
					(100,0.279)
					(150,0.424)
					(200,0.633)
					(250,0.779)
					(300,0.945)
					(350,1.106)
					(400,1.304)
					(450,1.642)
					(500,1.658)
				};
				\addlegendentry{\texttt{linear}}
				\addplot[MidnightBlue, mark=diamond*,mark indices=10] plot coordinates {
					(50,0.191)
					(100,0.49)
					(150,0.774)
					(200,0.964)
					(250,1.365)
					(300,1.54)
					(350,1.83)
					(400,2.535)
					(450,3.053)
					(500,3.355)
				};
				\addlegendentry{\texttt{random}}
				\addplot[ForestGreen, mark=square*,mark indices=10] plot coordinates {
					(50,0.08)
					(100,0.1575)
					(150,0.24)
					(200,0.3475)
					(250,0.425)
					(300,0.485)
					(350,0.5625)
					(400,0.6525)
					(450,0.755)
					(500,0.8325)
				};
				\addlegendentry{\texttt{tree}}
			\end{axis}
		\end{tikzpicture}
		}

	\caption{Runtimes for the computing responsibility values for the three model families \texttt{linear}, \texttt{random} and \texttt{tree} 
	for increasing numbers of actors and states.}
	\label{fig:performance}
\end{figure}

\newcommand{\thsep}{\kern0.15em}
Figure~\ref{fig:performance_actors} shows the computation time for constant model 
size $n=100\thsep{}000$ and varying numbers of actors. Performance differs between the models,
but for all three models it becomes evident that the runtime grows exponentially 
in the number of actors. 

Figure~\ref{fig:performance_states} shows the computation time for varying model sizes 
from $n=50\thsep{}000$ to $n=500\thsep{}000$. The number of actors is $7$ in every case. Due to the
overhead costs for exporting and re-importing, we did not scale our experiments 
beyond this point. For  $n=500\thsep{}000$, this step took four minutes already. One can observe
a strong linear correlation between the size of the transition system and the runtime.

Our results show that by choosing a coarse actor signature, it is possible to 
analyse large models with hundreds of thousands of states. Currently, transferring 
the data between model checker and the implementation creates a bottleneck, but this 
can be avoided by integrating responsibility directly into a model checker.

%% file: 07-conclusion.tex
\section{Conclusion}
\label{sec:conclusion}

This article builds on previous work on 
responsibility notions for (sets of) states on the satisfaction or violation of safety properties from the forward
\cite{ForwardResponsibility} and backward \cite{BackwardResponsibility} 
perspective. 
The focus here was on the view of actors and natural definitions of actor sets
that exploit structural system features (like modules or actions) 
and can be derived from the syntax of reactive programs.
Module-based actors are constructed by adding a scheduler module, value-based actors group states together in which one -- or multiple -- distinguished variables have the same value and action-based actor groups rely on an action-separation technique.
The experimental results carried out with a prototypical implementation served to provide insights in the scalability and how the presented responsibility notions can be used for debugging purposes.

\textbf{Future work.} The action-separation technique presented for action-based actors can be explored further. For example, it could be useful to build finer-grained actors that represent actions in individual states, instead of just considering actions globally. This would enable ascribing responsibility to individual transitions in the transition system.

While the focus of this article is on safety properties,
analogous definitions for the responsibility of actors on the satisfaction or violation of other properties, e.g., reachability, Büchi or LTL-formulas, can be provided \cite{ForwardResponsibility}.  Some of concepts presented in Section \ref{sec:reactive_modules} would need to be adapted as e.g. the soundness of the scheduler and action separation transformation (Lemma \ref{lemma:soundness-scheduler} and Lemma \ref{lemma:soundness-action}) does not hold when switching from safety to LTL properties.
 
The complexity results show that an efficient algorithm is unlikely to exist. However, there might be branch-and-bound techniques and other heuristics for finding switching pairs that provide a speed-up over the current implementation, which na\"ively enumerates all coalitions.